\documentclass[twocolumn,showpacs,preprintnumbers,amsmath,amssymb,prl]{revtex4}
\usepackage{graphicx}% Include figure files
\usepackage{dcolumn}% Align table columns on decimal point
\usepackage{bm}% bold math

\begin{document}

\preprint{APS/123-QED}

\title{The Stabilized Kuramoto-Sivashinsky Equation: A Useful Model For Secondary Instabilities and Related Dynamics of Experimental One-Dimensional Cellular Flows}% Force line breaks with \\

\author{P. Brunet$^{1,2}$}
%\altaffiliation[Present address: ]{University of Bristol - Department of Mathematics, University Walk, BS8 1TW Bristol (UK) \\}%Lines break automatically or can be forced with \\
\email{p.brunet@bristol.ac.uk}

\affiliation{%
$^{1}$ KTH Mechanics - SE - 10044 Stockholm, Sweden\\
$^{2}$ School of Mathematics, University of Bristol, University Walk, Bristol BS8 1TW, United Kingdom}%

%\affiliation{

\date{\today}% It is always \today, today,

\begin{abstract}

We report numerical simulations of one-dimensional cellular solutions of the stabilized Kuramoto-Sivashinsky (SKS) equation. This equation offers a range of generic behavior in pattern-forming instabilities of moving interfaces, like a host of secondary instabilities or transition toward disorder. We compare some of these collective behaviors to those observed in experiments. In particular destabilization scenarios of bifurcated states are studied in a spatially semi-extended situation, which is common in realistic patterns, but was barely explored so far.

\end{abstract}

\maketitle

An interface that is driven out-of-equilibrium frequently develops a patterned structure, characterized by a spatially periodic array of identical cells. This pattern can in turn show a set of various secondary instabilities, until a possible transition to disorder \cite{CrossHohenberg93}. Such a scenario has been encountered in various experiments, such as directional solidification \cite{Flesselles91,Ginibre97}, directional viscous fingering \cite{Rabaud90,Rabaud92,Debruyn93}, Taylor-Dean flow \cite{Riecke96}, a locally heated thin layer of liquid \cite{Burguete03} or an array of falling liquid columns \cite{Limat98,Brunet01,Brunet03,Brunet04,Brunet05}. 

Coullet and Iooss \cite{CoulletIooss90} have proposed a generic model, based on broken symmetries that predicts ten secondary instabilities from a primary static periodic cellular structure. This model reproduced successfully many features of secondary modes associated to broken symmetries on the primary pattern, for instance parity-broken (PB) domains of drifting cells \cite{C3G91} or vacillating-breathing (VB) mode leading to out-of-phase oscillations. However, the model was built under assumptions of slow-varying space-phase variables and thus remained valid only close to secondary thresholds. Gil \cite{Gil99,Gil00} has recently built an extension of this model, that includes possible phase-mismatch between the primary static state and the bifurcated one. Therefore, Gil's model could reproduce some far-from-secondary-threshold behaviors, such as oscillating patches left behind a propagative domain, black solitons or spatiotemporal disorder.

Alternative approaches using partial differential equations or cellular automata have been proposed, where the possible dynamical modes are not explicitly introduced in the model, but rather appear via the unstable dynamics of cellular solutions. An example is the stabilized Kuramoto-Sivashinsky (SKS) equation, which is investigated numerically in this paper. In its non-stabilized form, this equation was first  built to reproduce some general phenomena in falling film on inclined substrates \cite{Kuramoto78} or in flame-front instabilities \cite{Sivashinsky77}. The SKS equation is the following:

\begin{equation}
\frac{\partial f}{\partial t} = - \alpha f + \left( \frac{\partial f}{\partial x} \right)^2 - \frac{\partial^2 f}{\partial x^2} - \frac{\partial^4 f}{\partial x^4}  
\label{eq:ksa}
\end{equation}

The term $-\alpha f$ represents the damping term. It has been shown \cite{Misbah94} that this equation was one of the simplest to capture ubiquitous features of pattern-forming instabilities in interfacial growing fronts. This equation is also known to exhibit spatio-temporal intermittency, i.e. co-existence of laminar domains and turbulent patches for a large number of cells \cite{ChateManneville87}, while secondary bifurcations have been found as well \cite{Misbah94}, albeit for a small number of cells (up to 3). However, very few studies focussed on secondary instabilities for an intermediate number of cells (typically a few tens). This is the condition under which most experimental interfacial patterns are investigated, and it is expected that collective behaviors with both spatial and temporal significance show up. 

In this paper we numerically investigate tertiary bifurcations, i.e. destabilization scenarios on secondary dynamical states. These secondary states are themselves results of the destabilization of a primary static periodic structure. With this semi-constrained (or semi-extended) geometry, the pattern is large enough to show collective behaviors, and small enough to allow for the tracking of the motion and the shape of a single cell. We aim to find a generic comprehensive model for complex states that appear to be due to non-trivial mode coupling or finite-size effects: (a) oscillating wakes behind a propagative domain, (b) an amplitude hole corresponding to a phase jump in an extended oscillatory state, (c) oscillations superimposed on a state of drifting cells before its rupture.
We show that these tertiary states, amongst others, can be reproduced by the SKS equation, demonstrating its relevance for phenomena beyond the threshold of secondary instabilities. This paper is divided into two parts: we give a short description of the numerical method, followed by a set of results in the form of spatio-temporal diagrams. Then we comment on similarities and differences with experiments.

\paragraph{Results}

The resolution of eq. (\ref{eq:ksa}) is carried out by a pseudo-spectral method. The space derivatives are calculated in Fourier space: a multiplication of the vector of the Fourier coefficients, by the vector of the corresponding wave-numbers times the complex imaginary unit $i$, gives the first space derivative. Any $n^{th}$ space derivative is calculated via the same type of multiplication that is repeated $n$ times. The time derivative of $f$ is then evaluated by a finite difference method. The choice of a small time-step (typically around 10$^{-3}$) is suitable in order to avoid convergence problems. An implicit scheme has been used as well, but is not necessary here: the cellular solutions are smooth enough for a simple explicit scheme to work. We employ periodic boundary conditions and the initial conditions are fixed through the number of cells $N_c$, using about 25 mesh points per cell.

Inspired by the method employed in \cite{Misbah94}, we prescribe initial conditions as combinations of sinusoidal functions in order to trigger secondary instabilities from a primary periodic, static pattern. Most of the initials conditions consist of a single wave-number $k$ plus random perturbations (typical magnitude 1/100). Some states needed specific initial conditions, like for example the PB drifting cells, which are obtained from a combination: $f_0(x) = \sin (k x) + a \sin (2 k x + \phi_0)$. The phase shift is arbitrarily chosen at 0.5, and the amplitude $a$ is chosen to be equal to 0.5. These two quantities can take values in a certain range (0.3 to 2 for the absolute value of $\phi_0$, 0.25 to 1 for $a$) without changing the phenomenon qualitatively, but they slightly influence the kinetic properties of the selected states. As the purpose of this study is to seek for destabilization scenarios of secondary bifurcated states, we opt for a set of parameters ($k$, $\alpha$ and $N_C$) such that the system is close to the boundary of existence of the bifurcated states. To compute the temporal evolution of a localized domain of PB cells, we chose the following initial condition: $f_0(x) = \sin \left( k x (1+ \frac{\theta}{2} (1+ \tanh (x_{\text{lim}} - x)) \right) + a \sin \left( 2 k x + \phi_0) \right) \times \frac{1}{2} (\tanh (x_{\text{lim}} - x) +1) $. Then between $x$ = 0 and $x_{\text{lim}}$, the cells are asymmetrical and have a larger wavelength, which corresponds to the domain of stability of drifting cells \cite{Misbah94}.

\begin{figure}
\begin{center}
\includegraphics[scale=0.48]{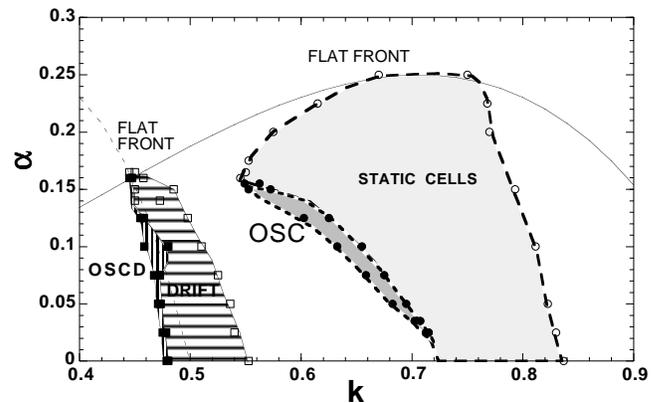}
\caption{Stability diagram of the primary and secondary states in the SKS equation. The domains for static cells, oscillating cells (OSC) drifting cells (DRIFT) and oscillating-drifting cells (OSCD) are bounded respectively by open circles, dark circles, open squares and dark squares. In empty domains, a given state is unstable. The full line stands for the neutral curve of the mode of wave number $k$, and the dashed line stands for the neutral curve of the mode $2k$.}
\label{fig:busse_balloon}
\end{center}
\end{figure}

Figure \ref{fig:busse_balloon} gives a cartography of the main states obtained by varying both $\alpha$ and $k$. The stability of each state is checked by runs of relatively long duration time of one hundred (corresponding to 10$^{5}$ time steps): if the state is not broken, it is considered as stable. The symbols stand for the domain boundaries: when these are crossed, the initial state undergoes a transition to another one (sometimes disordered). One of the main reason for the break-up is the occurrence of the Eckhaus instability \cite{Misbah94}, which delimitates the domain for static cells. The break-up of drifting states coincides well with the neutral curve of the mode of wavelength $2k$. The OSC regime is stable within a much narrower strip than the one predicted by the linear stability analysis \cite{Misbah94}. At large $\alpha$, both drifting and static states undergo a transition to a flat front and the cellular pattern vanishes: it corresponds to the neutral curve of the mode $k$.

\begin{figure}
\begin{center}
(a)\includegraphics[scale=0.26]{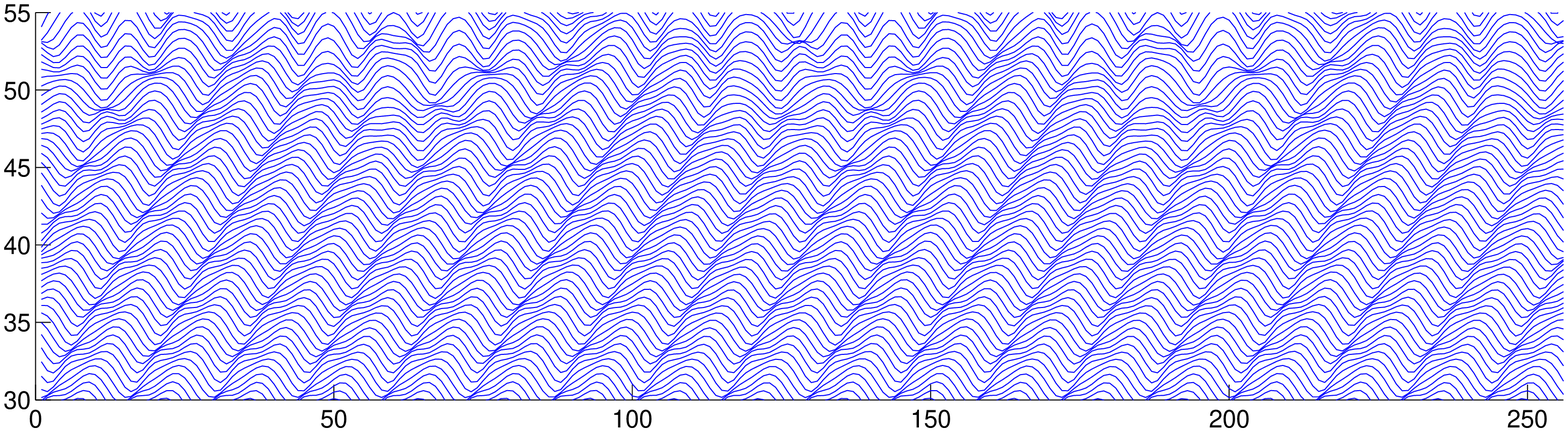}
(b)\includegraphics[scale=0.26]{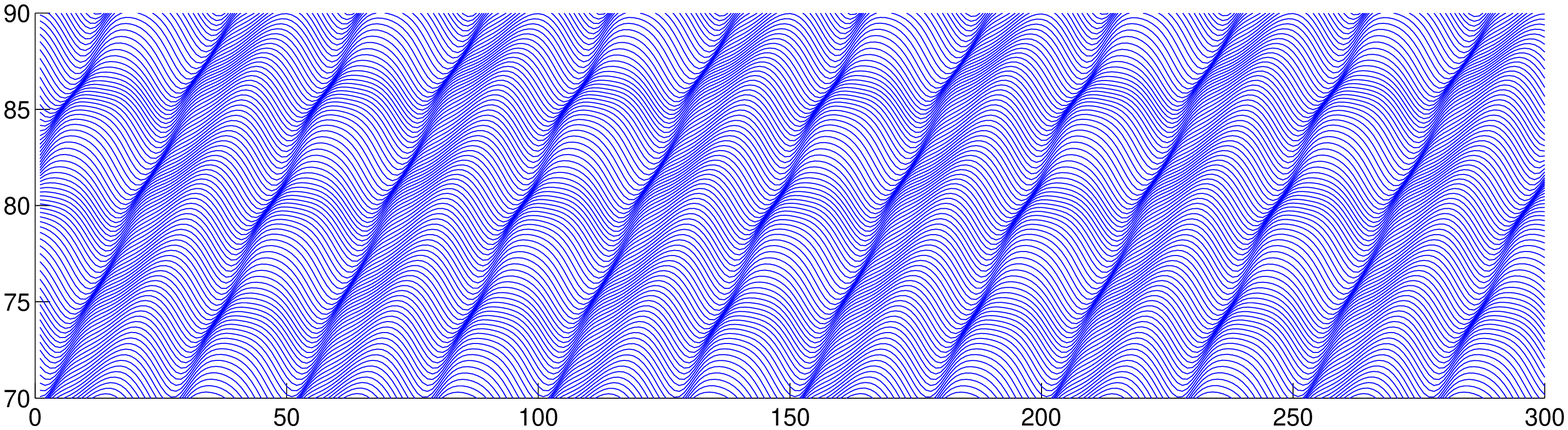}
(c)\includegraphics[scale=0.25]{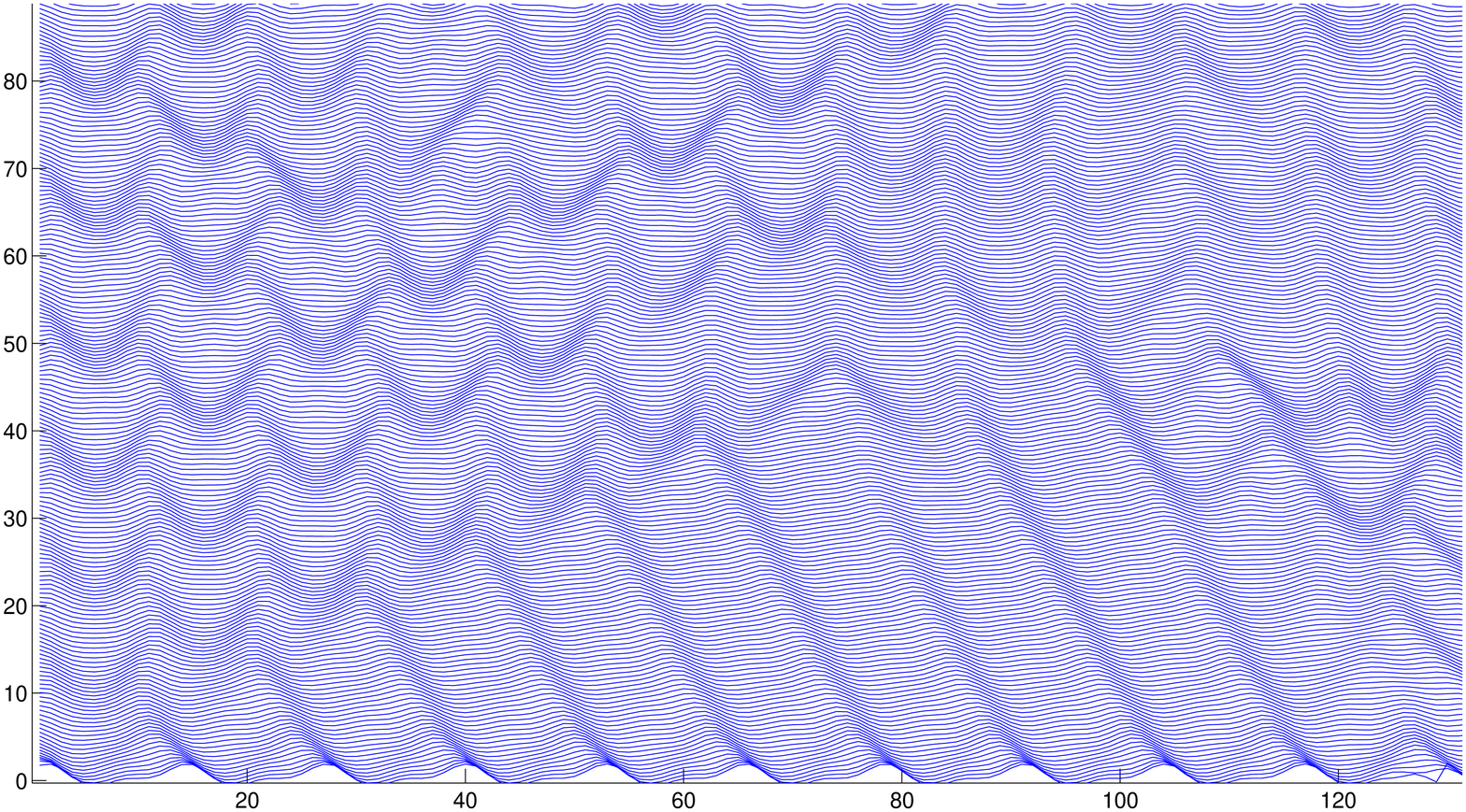}
%(d)\includegraphics[scale=0.25]{./chaos.eps}
\caption{Examples of complex dynamics beyond a secondary instability. (a) A global drift that undergoes in-phase oscillations as a first stage to the break-up towards spatiotemporal disorder ($k$=0.47, $\alpha$=0.095, $N_c$=18). (b) A global drift that undergoes a bifurcation to out-of-phase oscillations ($k$=0.53, $\alpha$=0.04, $N_c$=12). (c) Oscillating wake left behind a domain of drifting cells (extract) ($\alpha$= 0.15, $N_c$=18).}
\label{fig:ksachaos}
\end{center}
\end{figure}

Typical spatiotemporal diagrams are depicted on fig. \ref{fig:ksachaos}, and reproduce complex collective behaviors (a-c) reminiscent to various experiments of patterned interfaces. Time runs vertically from bottom to top. These diagrams are obtained by plotting $f(x,t)$ every n$^{th}$ time-step, added to a vertical shift proportional to the value of time. The case (a) represents a state of drifting cells, initially traveling at constant speed, which undergo an oscillatory instability in a second step, to ultimately break-up and enter a disordered regime. The case (b) is similar to (a), except that each cell oscillate out-of-phase of its nearest neighbors. These oscillations are likely to get amplified and lead to spatiotemporal disorder, like in case (a). The case (c) shows a local domain of drifting cells, propagating opposite of drift, and leaving oscillatory patches behind its trailing edge. 

\begin{figure}
\begin{center}
\includegraphics[scale=0.30]{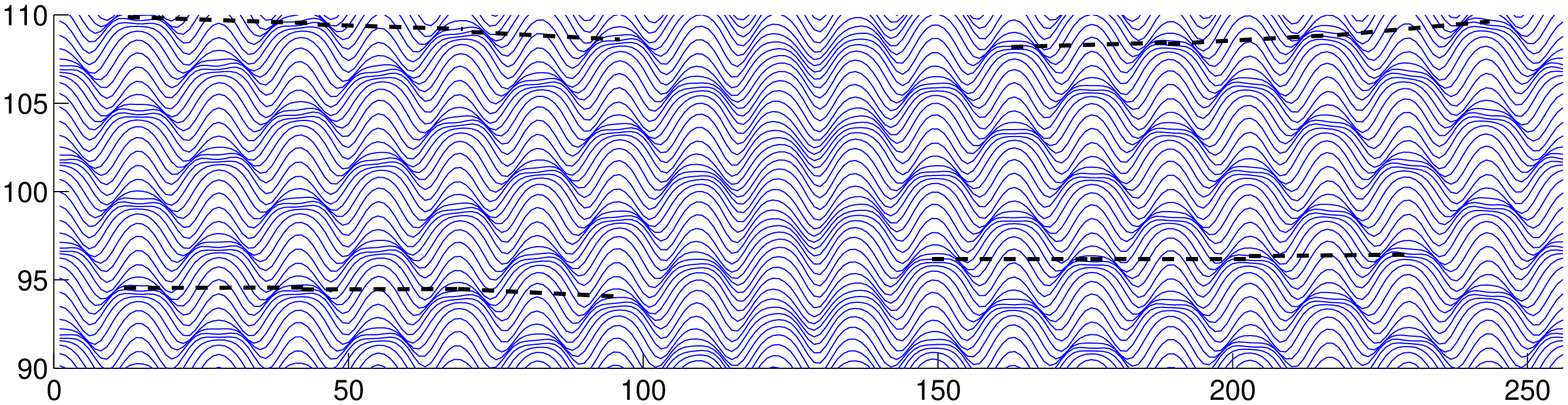}
\includegraphics[scale=0.28]{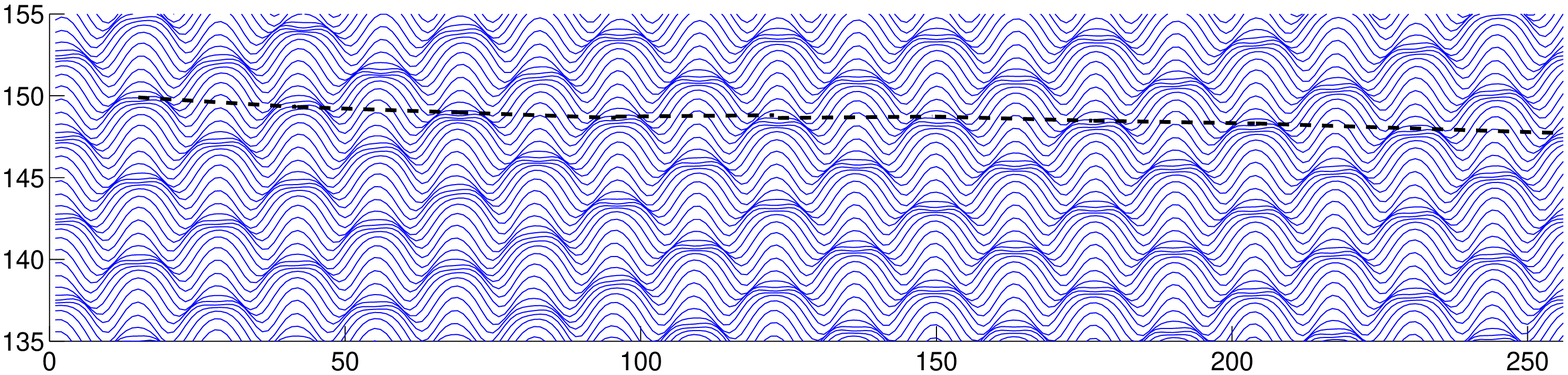}
\caption{Imperfect extended oscillatory, period-doubling state, caused by an odd number of cells and periodic boundaries. Top: localization of an amplitude hole. Bottom: progressive phase-shift. Both obtained with: $k$=0.64, $\alpha$=0.1, $N_c$=19.}
\label{fig:ksaoscil}
\end{center}
\end{figure}

A second kind of behavior concerns phase imperfections in extended oscillatory regimes, fig. \ref{fig:ksaoscil}. These diagrams are obtained when an oscillatory state develops and exhibits a so-called 'optical mode' (by analogy to eigenmodes in phonons) i.e. neighbor cells oscillating out of phase. While such a state can be almost perfectly homogeneous if the number of columns is even, it will necessarily adapt to a non-homogeneous spatial phase for an odd number of columns. These diagrams show that the phase imperfection can be twofold: it can either be sharply localized (top diagram on fig. \ref{fig:ksaoscil}) or evenly dispatched on the pattern (bottom diagram). In this second case, the dashed line describes the iso-phase of the double-period state. The first case exhibits an amplitude hole, i.e. the amplitude of oscillations vanishes in the vicinity of the phase imperfection, enabling the pattern to have a sharp phase jump of $\pi$.

\paragraph{Comparisons with experiments - Discussion} 

\begin{figure}
\begin{center}
\includegraphics[scale=0.33]{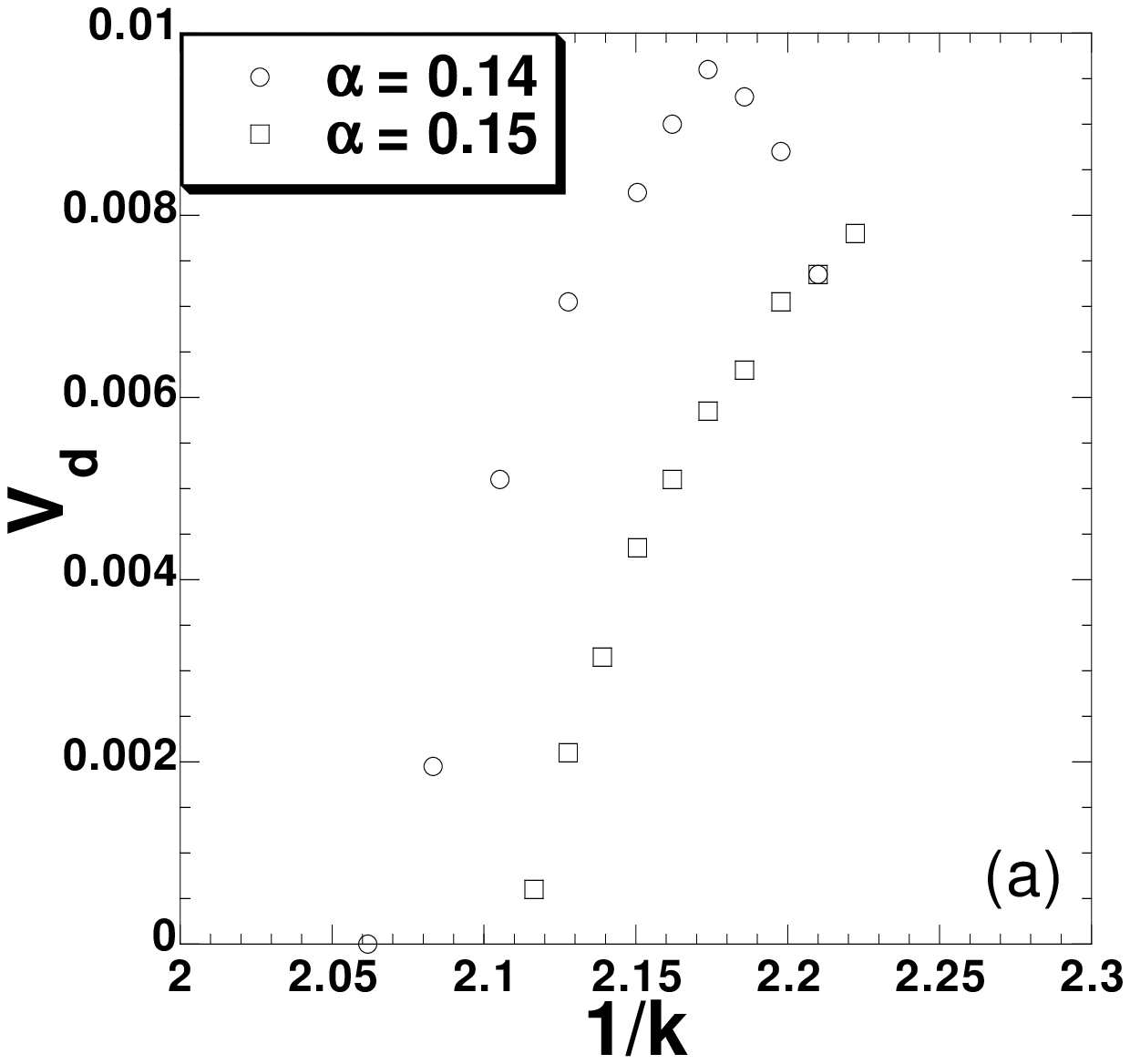}
\includegraphics[scale=0.33]{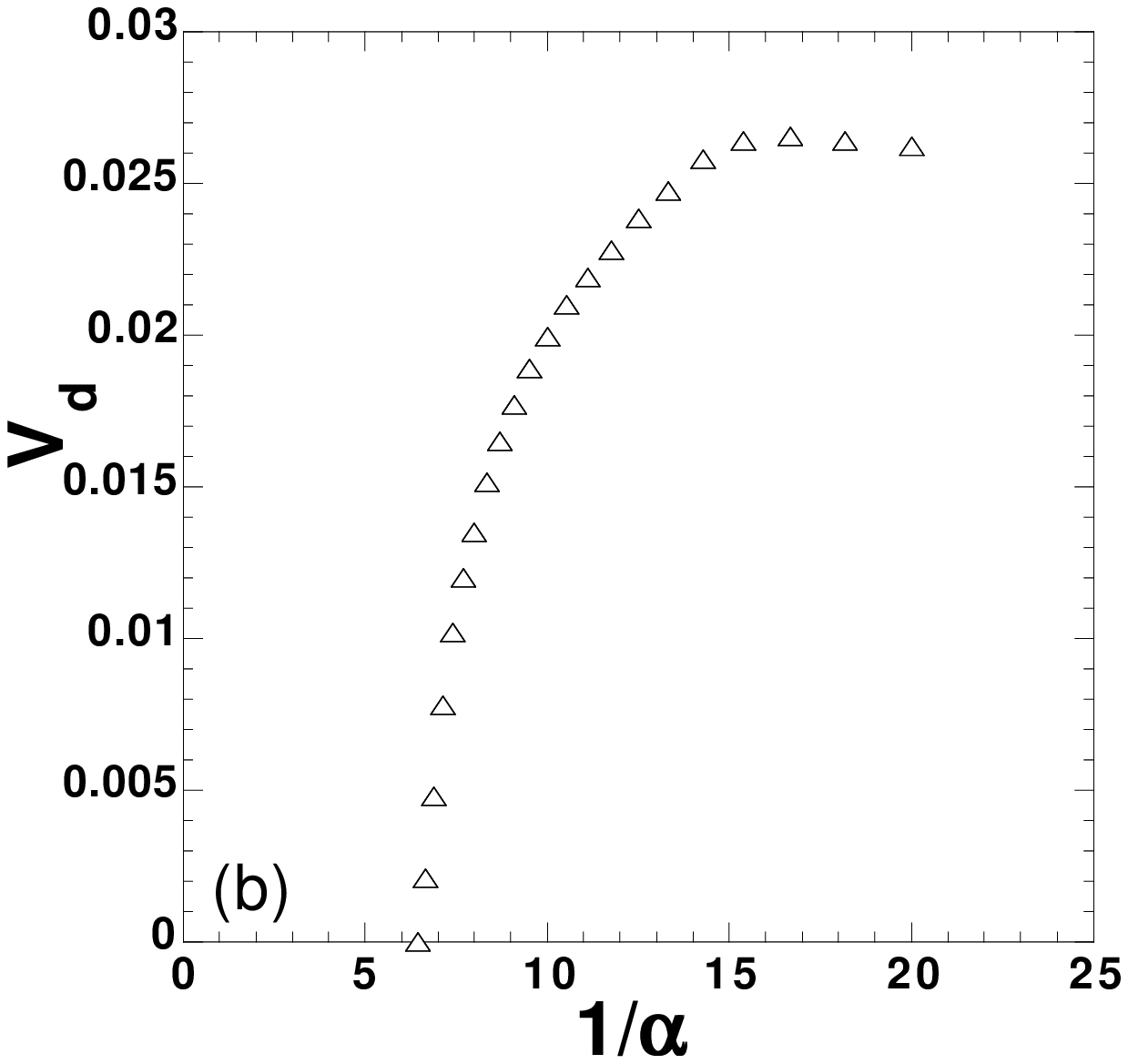}
\includegraphics[scale=0.33]{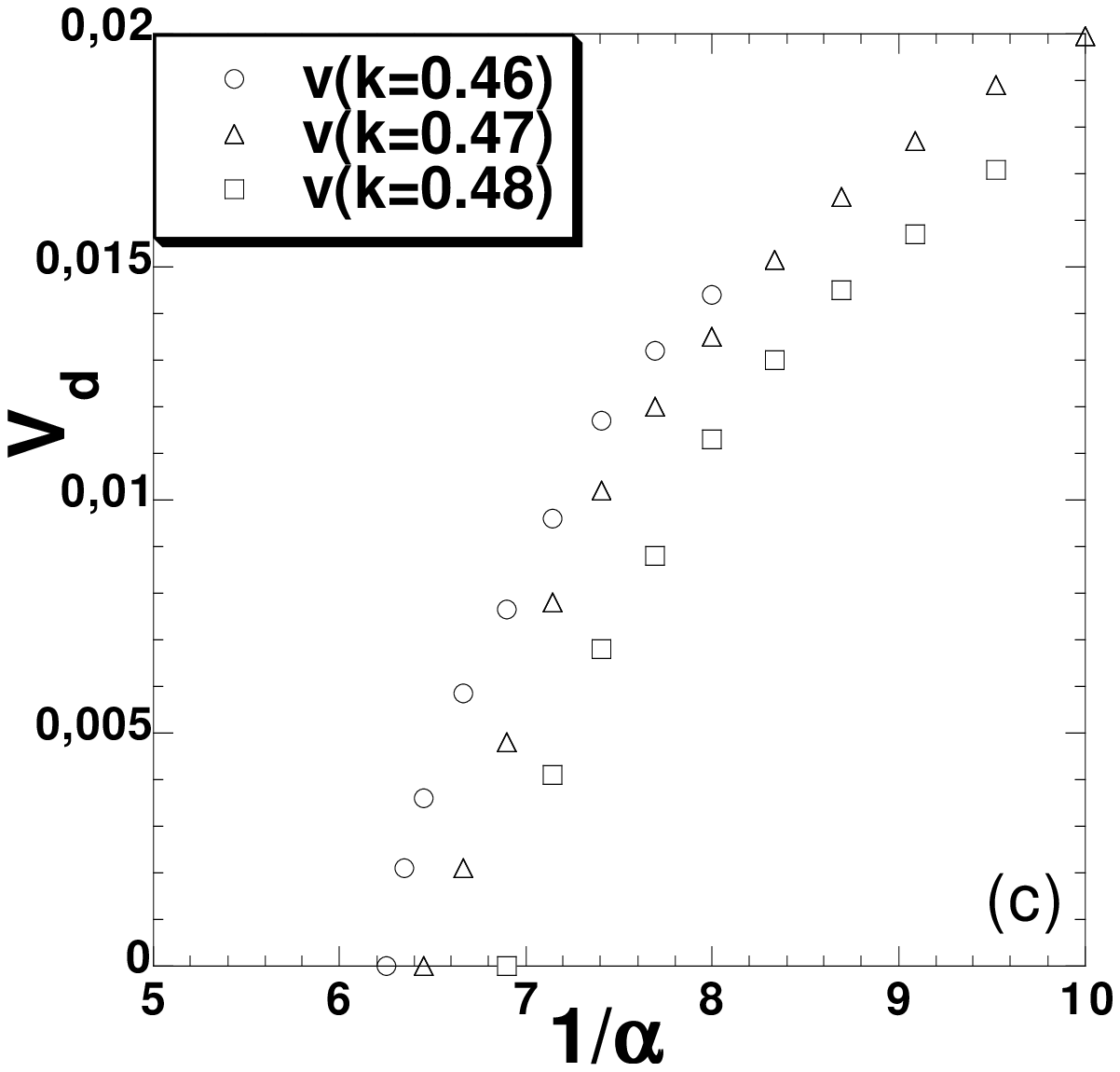}
\includegraphics[scale=0.33]{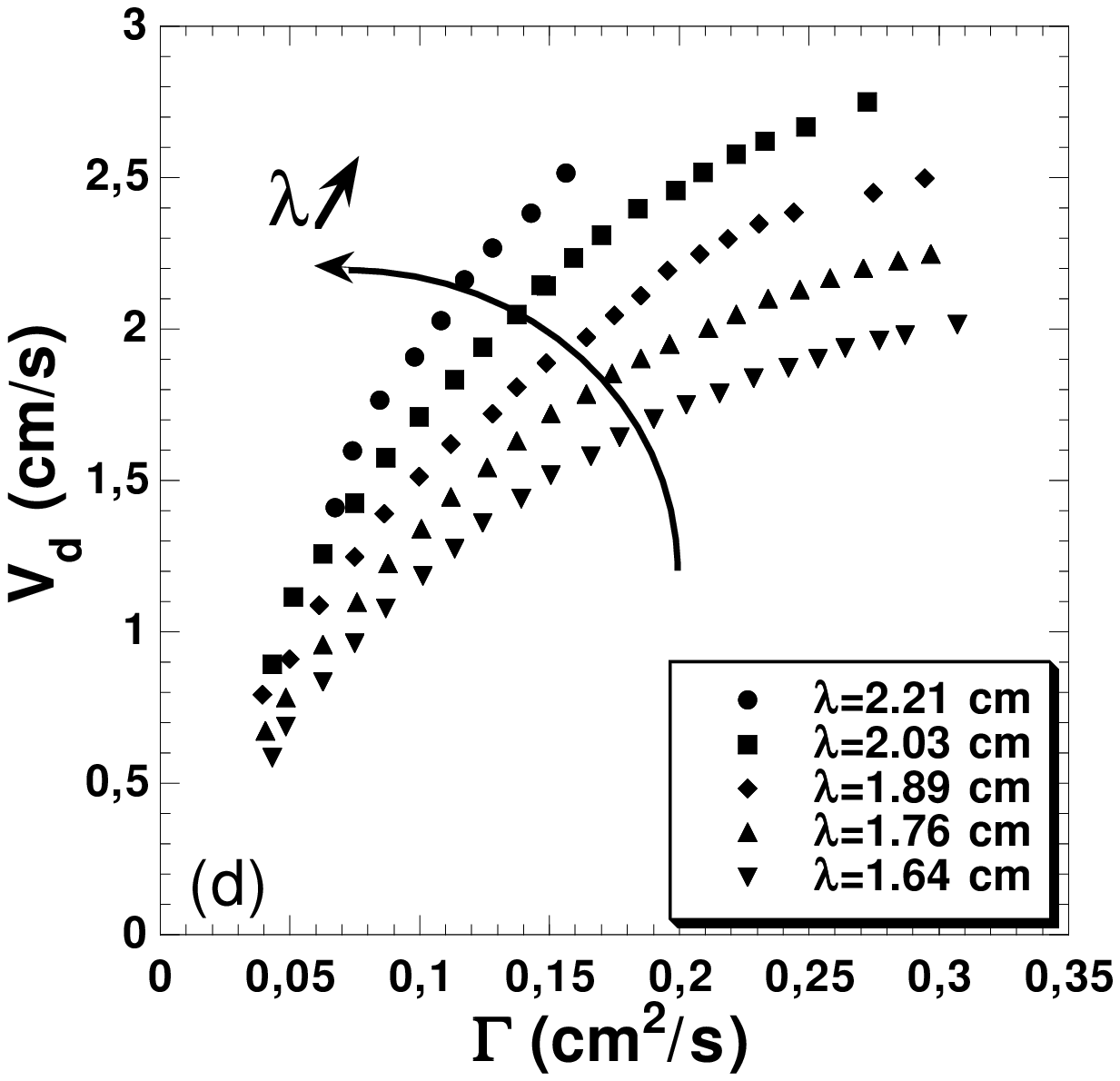}
\caption{Measurements of drifting velocities in the SKS equation (a,b,c) and in the experimental pattern of liquid jets (d). (a) Versus $1/k$ for two values of $\alpha$. (b) Versus $1/\alpha$ for $k$=0.47. (c) Versus $1/\alpha$ for various $k$ (close-up around threshold). (d) Drift velocities versus flow-rate, for different wavelengths (silicon oil of viscosity $\eta$=100 cP), }
\label{fig:vd}
\end{center}
\end{figure}

We now argue that the features depicted above reproduce destabilization scenarios observed in several experiments. We start with a presentation of quantitative data on the PB drifting cells. Figures \ref{fig:vd} show general trends of the drift speed $V_d$, measured on a globally extended drifting state in both the SKS equation (a-c) and a pattern of falling jets (d) \cite{Brunet01,Brunet05}. The distance along $x$ is chosen such as the length of a cell equals one. Figure \ref{fig:vd}-a shows the dependency of $V_d$ versus $1/k$, for two values of $\alpha$. Figures \ref{fig:vd}-b and -c show $V_d$ versus $1/\alpha$: the general trend is a sharp increase of the speed just above a threshold value for $\alpha$. Let us also recall that the drift velocity of PB cells in the pattern of columns ($V_d$) increases as the square-root of $\Gamma$ (the flow-rate per unit length), see fig. \ref{fig:vd}-d. If one tries to get a similar relationship for the SKS equation, it turns out that identifying the control parameter with $1/\alpha$, is the most relevant choice (compare figs. \ref{fig:vd}-b, \ref{fig:vd}-c and \ref{fig:vd}-d). In the pattern of columns, a similar dependance on $k$ was found: the drift speed increases with $1/k$ for most of the conditions (fig. \ref{fig:vd}-a). In other experiments, the identification of $1/\alpha$ with the control parameter of the PB bifurcation is also straightforward: in the printer's instability, it is the rotation speed of the internal cylinder \cite{Debruyn93} and in directional solidification it is a combination of the wavelength $\lambda$ and the pulling velocity $V$ that reads: $\lambda V^2$.

\begin{figure}
\begin{center}
(a)\includegraphics[scale=0.35]{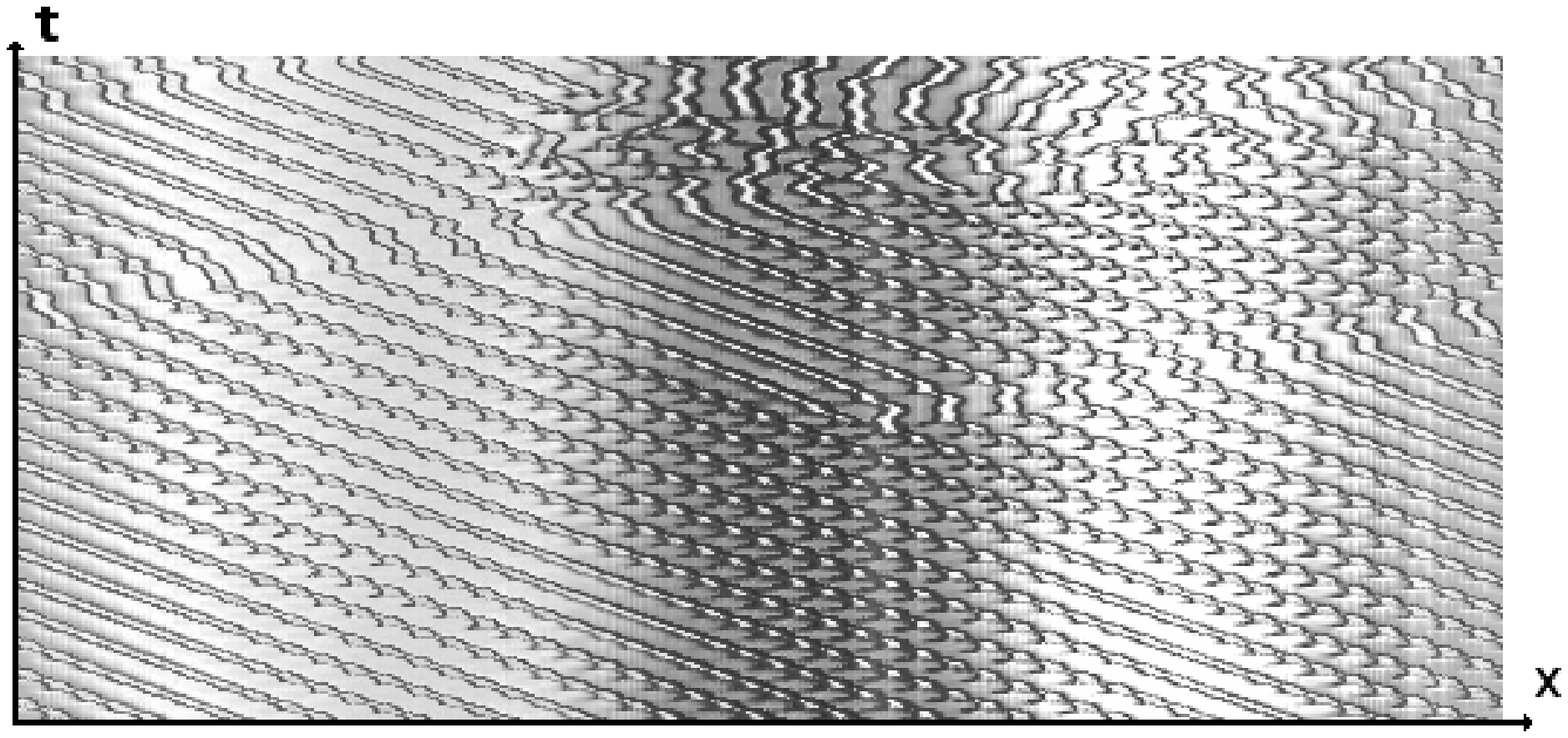}
(b)\includegraphics[scale=0.35]{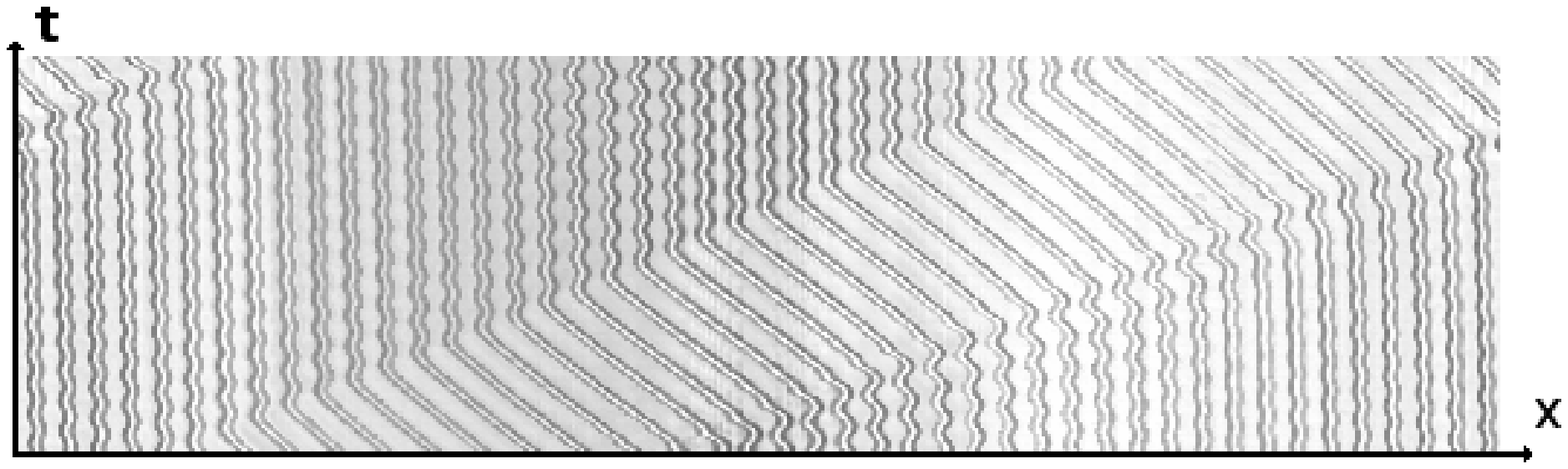}
(c)\includegraphics[scale=0.35]{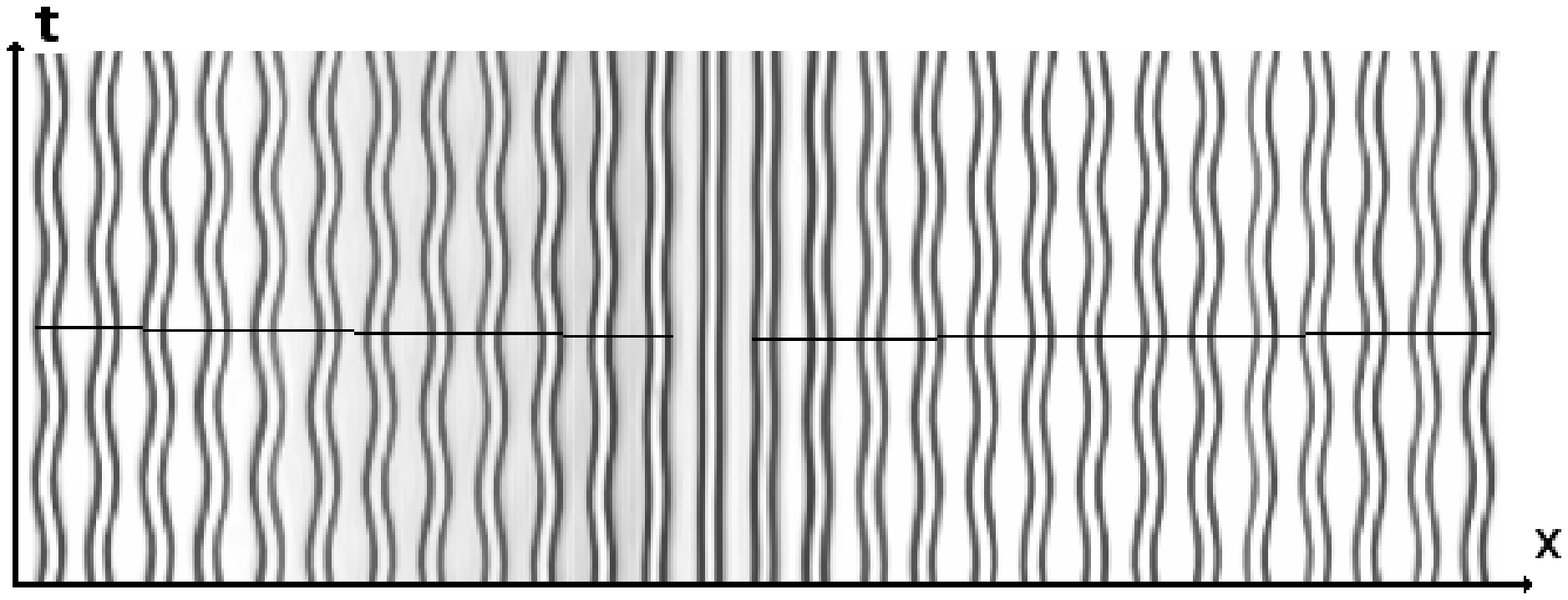}
(d)\includegraphics[scale=0.35]{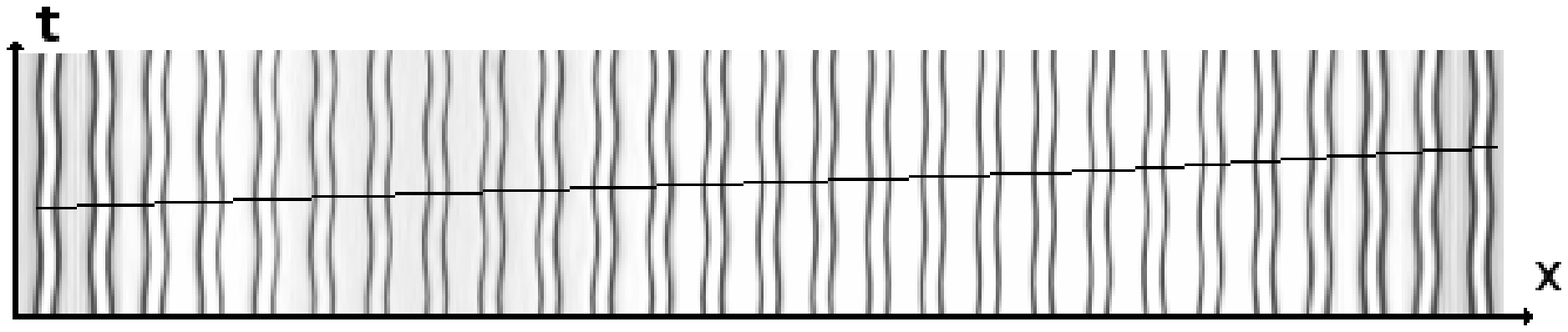}
\caption{Various collective behaviors in the pattern of liquid columns (viscosity $\eta$=100 cP), reproduced by the SKS equation. (a) Oscillations superimposed on a state of global drifting cells, leading ultimately to disorder. (b) An oscillating wake in the trailing edge of a propagating domain. (c) A phase defect localized in an oscillatory state. (d) Progressive phase-shift in an oscillatory state.}
\label{fig:spatioexp}
\end{center}
\end{figure}

The dynamics depicted in figures \ref{fig:ksachaos} are commonly observed in experiments. Case (a) is reminiscent to various experiments \cite{Ginibre97,Riecke96,Brunet05}. It can either appear under temporal modulations of the control parameter \cite{Riecke96} or simply when the control parameter is increased sufficiently beyond the threshold of drifting cells \cite{Ginibre97,Brunet05}. Case (b) has only been observed in directional solidification and referred to as T-$2\lambda$ O in \cite{Ginibre97}. Case (c) is observed in both directional viscous fingering \cite{Rabaud92} and an array of liquid columns \cite{Brunet03,Brunet05}. It shows the link between the parity-breaking and vacillating-breathing modes: in this case the relaxation at the rear wall of the propagative dilation wave (drifting cells are dilated compared to static cells) leads to oscillations at the trailing edge of the domain. These oscillations are the local counterparts of the ones in fig. \ref{fig:ksaoscil}. Two of these destabilizing behaviors are reproduced on figs. \ref{fig:spatioexp}-a and b for the pattern of liquid columns \cite{Brunet03,Brunet05}.

We now examine the phase imperfections in an oscillatory state (figs. \ref{fig:ksaoscil}). The two types of phase imperfections on an extended oscillating state have been observed in the pattern of columns \cite{Brunet05} for an odd number of columns (whereas an almost homogeneous oscillatory state is observed for an even number of columns), as shown on figs. \ref{fig:spatioexp}-c and d. It should be noted, however, that the SKS equation failed to reproduce a sustained propagating domain of PB cells. Such a domain is bound to shrink and to vanish after a while, at least in the explored parameter range. Thus probably, a state of PB cells remains stable only for an extended drifting state.

In conclusion, the presented results show that the SKS equation is able to reproduce a set of complex situations that occur for some secondary instabilities of pattern-forming experiments. We have chosen a semi-extended pattern (a few tens of cells), as this is frequently encountered in experiments, and we have taken parameters and initial conditions in order to trigger further destabilization of secondary bifurcated states. The fact that such realistic behaviors are reproduced by a simple generic equation like   (\ref{eq:ksa}), provides an interesting perspective for studying tertiary bifurcations. This includes a disordered regime with occurrences of phase defects (such defects appear on top of diagram fig. \ref{fig:ksachaos}-a), already reported in \cite{Misbah94} for a small number of cells and also obtained for a larger number of cells (a few tens) in our simulations. A comprehensive study of such a regime based on statistics of defect occurrences, would be of great interest.

\paragraph{Acknowledgments - } J. Hoepffner and L. Brandt are kindly acknowledged for their help in building the code. We thank J.H. Snoeijer for a critical reading of the manuscript.

%\newpage %Just because of unusual number of tables stacked at end
\bibliography{apssamp}% Produces the bibliography via BibTeX.

\end{document}